\documentclass [10pt]{article}

\usepackage{amsmath}
\usepackage{amsmath,amsfonts,latexsym}
\usepackage{amssymb}
\usepackage{calrsfs}

\newtheorem{lemma}{Lemma}
\newtheorem{theorem}{Theorem}

\newtheorem{proposition}{Proposition}
\newtheorem{corollary}{Corollary}

\usepackage{graphicx}
\graphicspath{%
    {converted_graphics/}
    {Z:/mydocs/BORKARpub/IEEE_Trans_Network/IEEEtran/NetSci&Eng_figures/converted_graphics/}
    {Z:/mydocs/BORKARpub/IEEE_Trans_Network/IEEEtran/NetSci&Eng_figures/}
    {Z:/mydocs/BORKARpub/IEEE_Trans_Network/IEEEtran/converted_graphics/}
    {Z:/mydocs/BORKARpub/SEA-14/llncs2e/Hunt_formerTNSE18/converted_graphics/}
    {Z:/mydocs/BORKARpub/SEA-14/llncs2e/OptimalSeeds/pngs-for-OptRec/}
    {pngs-for-OptRec/}
}
\begin{document}
%
%
%
\title{Using First Hitting Times to Find Sets that Maximize the Convergence Rate to Consensus}
%
%
\author{Fern Y. Hunt\\
Applied and Computational Mathematics Division\\
National Institute of Standards and Technology\\
Gaithersburg,Maryland 20899}        
\date{\today}          
\maketitle


%

\maketitle

\begin{abstract}
In a model of communication in a social network described by a simple consensus model, we pose the problem of finding a subset of nodes with given cardinality and fixed consensus values that enable the fastest convergence rate to equilibrium of the values of the remaining nodes. Given a network topology and a subset, called the stubborn nodes, the equilibrium exists and is a convex sum of the initial values of the stubborn nodes (\cite{GhaderiSrikant}, \cite{Pirani}, \cite{Como}).  The value at a non-stubborn node converges to its consensus value exponentially with a rate constant determined by the expected first hitting time of a random walker starting at the node and ending at the first stubborn node it visits.
In this paper, following the work of \cite{Borkar}, \cite{Clark} \cite{Yildiz} we will use the sum of the expected first hitting times to the stubborn nodes as an objective function for a minimization problem. Its solution is a set with the fastest convergence rate. 
We present a polynomial time method for obtaining approximate solutions of the optimization problem for fixed cardinality less than that of a reference vertex cover. Under the assumption that the transition matrix for the random walk is irreducible and reversible, we also obtain an upper bound for the expected first hitting time and therefore an upper bound on the rate of convergence to consensus, using results from the mixing theory of Markov chains. In addition to serving as a convenient screen for sets with fast convergence, the bound also reveals a new characterization of the convergence to consensus in terms of the bottle neck ratio of the Markov chain of the random walk and an accessibility measure expressed in terms the convex weights used in the consensus value.
%
%

\end{abstract}


%
%
%
\section{Introduction}\label{S:rwalk}
The DeGroot model of consensus formation among individuals or agents in a network has been the subject of intense research since its introduction in 1974 \cite{DeGroot}. In the simplest formulation, individual agents synchronously update their function value by averaging it with that of their neighbors. This process is repeated and the resulting function values converge to a single equilibrium thus modelling the evolution of agent opinions to consensus. More sophisticated variations have been studied e.g. allowing for asynchronous communication \cite{Boyd}, random and switching network topologies, and time varying averaging schemes
\cite{Tsit2}. There has also been work on other models of communication e.g. the voter model \cite{Acemoglu} \cite{Yildiz}.
In this paper, we will focus on the case where a subset of agents retain their initial values throughout the process. In the area of opinion dynamics, they are referred to as "stubborn agents". As shown e.g. in \cite{GhaderiSrikant},\cite{Como}, the equilibrium or consensus value of each node is a convex combination of the initial values of the stubborn nodes.  

In our discussion, the dynamics of the process will not play a role here since it will always be assumed (for simplicity) that the equilibrium value is the same constant for each agent.  Rather, we will discuss the problem of finding sets of stubborn agents that lead to the fastest convergence to equilibrium. Here as in recent work of Hunt \cite{Hunt1} and \cite{Hunt2}, and following earlier work of Borkar \cite{Borkar}, we formulate an optimization problem based on a random walk on the underlying graph associated with the averaging matrix. We use the fact that the convergence time is asymptotically equivalent to the expected first hitting time of the random walk to the nodes representing the stubborn agents \cite{Borkar},\cite{AldousFill}. The search for stubborn agents of fixed cardinality is then posed in terms of finding a set for which the first hitting time is smallest.
 A very similar problem arises in the design and control of leader-follower systems using distributed protocols see e.g. Clark et al, \cite{Clark} , \cite{Clark512}. 
A search for solutions of these optimization problems involve an exponential number of computations so polynomial time algorithms for approximate solutions are sought. In both situations, this is possible because the respective objective functions are submodular (or supermodular). Supermodularity of the first hitting time was established by Borkar \cite{Borkar} and later in the work of Clark et al \cite{Clark512} and Hunt \cite{Hunt1}. 
   
   Here our purpose is to further an understanding of the graphical structure of optimal and highly optimal sets in order to develop algorithms that can utilize these features. As a consequence, we may eventually identify classes of graphs for which we can efficiently solve or approximate the fast consensus problem and related node selection problems.
We begin with properties that can be deduced by using the expected first hitting time as an objective function. For the uniform random walk (with no self loops) it was shown \cite{Hunt1} that a vertex cover is an optimal set for its cardinality as is any superset of it. This is true in general for random walks with no self loops and for lazy versions of these walks. Since a vertex cover (not necessarily minimal) is easy to compute it suffices to restrict the problem to finding an approximately optimal set of cardinality smaller than that of some fixed vertex cover. Further given a vertex cover, all sets of smaller cardinality can be ranked relative to it: the smaller the objective function, the higher the rank. We present a polynomial time generalization of the classic greedy algorithm that begins with a starter set of optimal and nearly optimal sets rather than the best singleton set. The precise definition of a nearly optimal set is given in section\ref{S:optnopt}. For an $N$ vertex graph, and a starter set consisting of singletons, the computational effort is comparable to the greedy algorithm, $O(N*E)$ where $E$ is the computational effort to evaluate $F$ on a subset of $V$. For a starter set of two element sets it is $O(N^2*E)$. In \cite{Hunt1} our solution and the  greedy algorithm solution were compared using a submodular rank function. The supermodularity and non-increasing properties of our objective function were used to show that the rank of our solution is at least $(1-\frac{1}{e})$ times the rank of the optimal set is at least as good as the greedy solution. Moreover our computations show that our solution is often better and sufficient conditions that imply an improved lower bound on the rank can be found in section \ref{S:optnopt}
There are connections between the optimal and nearly optimal sets we discuss here and other sets of importance in network control. One dominant subsets are an example. A set is one dominant if every node in the complement is adjacent to a node that is in the set. A minimum dominating set is a dominant set of minimum cardinality. Recently,the importance of minimum dominating sets has been discussed theoretically and demonstrated in real networks \cite{Wuchty}, \cite{NachAkut1},\cite{NachAkut2}, including protein-protein interaction networks where research is suggesting that the nodes of these sets play very critical roles in network topology and function \cite{Wuchty2}. 
 We derive an upper bound for the expected first hitting time of  one dominant sets ( which include minimum dominant sets) and further show sufficient conditions for such a set to belong to a class of optimal and near optimal sets.

The convergence rate to consensus can also be characterized in terms of the leading eigenvalues of the matrix defining the dynamics. In the discrete time formulation it is the leading eigenvalue of the substochastic matrix corresponding to the non-stubborn nodes (see the work of Borkar \cite{Borkar}, Como and Fagnani \cite{Como}, Ghaderi and Srikant \cite{GhaderiSrikant}). In the continuous time case, the convergence rate is controlled by the smallest real eigenvalue of the submatrix of the Laplacian restricted to the non-stubborn nodes, i.e. the grounded Laplacian (\cite{Pirani2}). Upper  and lower bounds on this eigenvalue  were obtained by Pirani and Sundaram 
when the graph is undirected \cite{Pirani3}. Recently Xia and Cao \cite{Xia} established many similar lower and upper bounds for the directed case.The bounds on the eigenvalues obtained by these authors give insight into the graph topology of these sets.

In section \ref{S:Fupperbound} we derive upper bounds on the sum of the first hitting times to an arbitrary subset $A$ using recent results of Basu, Hermon and Peres in the mixing theory of Markov chains \cite{Basu}. No assumptions are made about the number of components in the complement of $A$, $A^{c}=V \setminus A$. The bound involves the bottleneck ratio of the boundary of $A^{c}$, the number of uncovered edges, i.e. edges with both endpoints in $A^{c}$, and lastly an accessibility measure; a weighted version of centrality measure that arises in a model of binary opinion dynamics in the presence of stubborn agents \cite{Yildiz}. The number of uncovered edges is a measure of the size of the components and the connectivity of $A^{c}$. The accessibility measure depends on the node and the set $A$ of stubborn agents. Roughly speaking it measures how accessible a node is to random walks that start at the boundary of $A^{c}$ that are eventually  absorbed on reaching $A$. This interaction of local and global topological properties of the graph with properties of random walks in the context of rate of convergence to consensus is new.
To illustrate the power of this approach a simplified version of the upper bound is used in section
\ref{S: Fupbound} to quickly screen for highly optimal sets.
\section{Notation}
 Since our paper concerns random walks on a graph we introduce the notation and concepts to be used. An exposition on Markov chain theory can be found in \cite{Kemeny}, while random walks on graphs are discussed in \cite{AldousFill}, \cite{Levin}. A graph is denoted by  $\mathcal{G}=(V, \mathcal{E})$, with $V$ the set of nodes and $\mathcal{E}$, its edges. An individual node is indicated by small Roman letters e.g. $i,j$. The graph is assumed to be strongly connected. Throughout our discussion a subset of nodes $A \subset V$, has a complementary set $A^{c}=V \setminus A$. A node in $A^{c}$ is called a boundary node if it is adjacent to a node in $A$. The notation $\cal{T}_{i}$ will be used to denote the connected component of $A^{c}$ that contains the node $i$. 

 The position at any time of a random walker traversing the nodes of a graph in discrete time steps can be described by a random process known as a Markov chain.  Here the states of the Markov chain are nodes of the graph and the state at any time is the node the random walker occupies at that time. Given an initial probability distribution on the nodes, the distribution after a single time step can be found in terms of the transition probability matrix $P$.  Using a vector $p$, let the $i$th component $p_i$   denote the probability of being at node $i$ at a fixed time step. The position at the next step is described by the probability distribution $p^{T}P$.  The ($i , j$)-th entry of $P$ , $P(i , j)$  is the conditional probability that random walker steps to $j$ given that the walker was at $i$ in the previous time step. Notions of conditional expectation are also used. The notation for the expected or mean value of a random variable $X$ is $\mathbb{E}[X]$, while the expected value $X$ given the random walker is initially at node $i$ is $\mathbb{E}_{i}[X]$.  More generally if the initial position of the walker is defined by a probability distribution $\mu$, then the expected value of $X$ given  $\mu$ is $\mathbb{E}_{\mu}[X]$.
When the Markov chain is irreducible and aperiodic, the distribution of possible positions of the random walker after a long time is described by the so-called stationary distribution, denoted by the row vector $\pi$. In addition we will assume the Markov chain is reversible so that $\pi (i)$$P( i , j)$=$\pi(j)$$P( j, i)$. The commonest example of such a process is the simple uniform random walk
whose transition probabilities are:
 
\begin{equation} \label{E: uniform}
P(i , j)=
\left\{
\begin{array}{ll}
\frac{1}{deg(i)},     &\text{ if $j\sim i$;} \\
0                              &\text{otherwise. }
\end{array}
\right.
\end{equation}
A node $i$ is \underline{adjacent} to $j$ iff $(i,j)$ or $(j,i) \in \mathcal{E}$. The notation for this relation is  $i\sim j$. The results of this paper apply to any walk on a directed graph with no self-loops that is described by a reversible Markov chain. Note that we are allowing weighted edges i.e. a non-symmetric adjacency matrix. The results on first hitting times and optimality can be extended to lazy versions of the random walk as well because the mean values of these random variables for the non-lazy case can be obtained by divideing by 2. A lazy version of the random walk governed by the transition matrix $P$ has a transition matrix
$(I+P)/2$. The same nodes and edges are used as in the original Markov chain and the stationary vector is the same for both chains. In this work $P(i,j)$ and $P(j,i)$ are assumed to be positive if $(i,j)$ is an edge and $i \ne j$. All the computations described in this paper used the uniform random walk or its lazy version.

In section \ref{S:setting} we will discuss the random variable that describes the first time a random walker visits a subset $A$, i.e. the first hitting time of  $A$, $T_{A}\triangleq\min\{t \geq 0: X_{t} \in A\}$ , where $X_{t}$
is the position of the walker at time $t$.
The dynamics of the random walk before time $T_A$ is described by the submatrix of $P$ obtained by crossing out the rows and columns of $P$ corresponding to the nodes of $A$. We denote this matrix by $P_A$ in section \ref{S:setting}. To describe the process of the random walk before the set $A$ is reached it is useful to count the number of times an edge $e$ is crossed. Let $n_{e}$ be the number of times a random walker crosses $e$ before time $T_{A}$.


\section{Problem Setting}\label{S:setting}
Although the rate of convergence to consensus  given a subset $A$ of stubborn or leader nodes is  determined by the leading eigenvalue of the sub-stochastic matrix $P_{A}$ , Clark et al \cite{Clark} have shown that the convergence rate of the error has an upper bound given by the first hitting time to $A$. This result can also be obtained by observing that convergence of the tail distribution of the first hitting time is asymptotically equivalent to an exponential with rate $1/(\max_{i}(E_{i}[T_A])$,  using for example \cite{Roch} (Chapter 3) and see also \cite{AldousFill}. Thus following \cite{Borkar} (equation 8), we seek a set that maximizes the convergence rate by minimizing $F$, the sum of the mean first hitting times of random walkers outside the set. Thus for the set $A$ let,
\begin{equation} 
F(A)=\sum_{i \notin A}E_{i}[T_{A}].
\end{equation}
If $A$ is an effective target set for the random walk then $F(A)$ will be small and this implies fast convergence to consensus. Thus subject to a constraint on the cardinality $K$ of the sets to be considered, we define the optimization problem,
\begin{equation} \label{E:opt}
\min_{A \subset V, \, |A|\leq K} F(A).  
\end{equation}
The problem as stated is quite difficult so approximations are sought that exploit properties of $F$.
We observe that if $A \subseteq B$, then $F(A) \geq F(B)$. Thus $F$ is a non-negative real valued non-increasing set function. Secondly it can be proved \cite{Borkar},\cite{Clark} that $F$ is supermodular. That is, $A\subseteq B \subseteq V$ and $j \in V$, $F(A)-F(A \cup j) \geq  F(B)-F(B \cup j)$. These properties also follow from an explicit formula
for $F(A)-F(A\cup j)$ derived in \cite{Hunt2}.
As a consequence $-F$ is submodular and when it is normalized (see section \ref{S:quality}), the greedy algorithm produces an approximation to the solution of problem \eqref{E:opt} that is within $(1-\frac{1}{e})$ of optimal
 \cite{NemWolFish}.    
 Given a set $A$, a standard result in Markov chain theory \cite{Kemeny} can be used to calculate $F(A)$. Let $h(i , A)=E_{i}[T_A]$.  Then it is the $i$th component of the vector solution $H$ of ,
\begin{equation} \label{E:lineq}
H=\mathbf{1}+P_{A}H. 
\end{equation}
Here $P_{A}$ is the matrix that results from removing from $P$, the rows and columns corresponding to the nodes in $A$ and $\mathbf{1}$ is the vector of $N-|A|$ ones. The value $F(A)$ is then computed by summing the components of $H$. 
From equation (\ref{E:lineq}) it immediately follows that if the Markov chain has no self loops any vertex cover is an optimal set, 
that is a solution of problem (\ref{E:opt}) for $K$=cardinality of the set. To see this recall that  $A$ is a vertex cover if every edge of $\mathcal{G}$ has at least one endpoint in $A$. This implies that for every
 $i \notin A$, $h( i , A)=1$. However equation (\ref{E:lineq}) implies that $h(i , A)\geq 1$, hence $F(A)$ is minimal. Since a superset of a vertex cover is also a vertex cover we see that if a vertex cover of cardinality $C$ is available, then the problem is solved for $K \geq C$ and only needs to be solved for $K <C$. A vertex cover (not necessarily minimal) can be easily computed. In particular, the vertices of a maximal match are a vertex cover and a straightforward algorithm of complexity $O(|\mathcal{E}|)$ produces a set no larger than twice the cardinality of the minimum vertex cover \cite{Cormen}. 
\section{Optimal and Near Optimal Sets}\label{S:optnopt}
There are two observations that are key to our approach to problem (\ref{E:opt}).  The first as discussed in section \ref{S:setting} is that vertex covers are optimal sets for their cardinality. The second is that a vertex cover often contains subsets that are optimal sets (see however \cite{Hunt1} for an example of a vertex cover that contains no optimal sets as proper subsets).  Thus given a vertex cover of cardinality $C$ and a  $K <C$,  the search for optimal sets is restricted to near optimal (or optimal) sets of cardinality $\leq K$. We expect that optimal sets of size $K$ much like vertex covers will contain subsets that are optimal or in some sense near optimal. Then generalizing the classic greedy approach, each set in a starter set (rather than the best singleton) is greedily extended in a stepwise fashion until a set of cardinality $K$ is reached. The offered approximation is the set with the smallest value of $F$ among all the extended sets. It remains then to make the term ``near optimal'' precise. We will rank all non-empty subsets $A\subset V : |A| \leq C$ with a ranking function $r$, defined as,
\begin{equation}\label {E:rank}
r(A)=\frac{F_{max}-F(A)}{F_{max}-F_{min}} 
\end{equation}
where $F_{max}=\max_{\emptyset \ne A\subseteq V, |A|\leq C}\large F(A)$, and $F_{min}$ is the corresponding minimum.

$F_{min}$ can be calculated by computing the $F$ value of  the vertex cover of cardinality $C$. We define $F_{max}$ to be the maximum value of $F$ among all one element subsets. We can assume that
$F_{max}\ne F_{min}$. Indeed if this was not the case, $F(A)$ would have the same value for any non-empty $A$ of cardinality less than $C$ and hence any such set would be a solution. The function $r$ captures the  notion of near optimal  desired. For example if $A$ is optimal and $|A|=C$ then $r(A)=1$, conversely for the worst performing set $r(A)=0$. We can now define a class of near optimal sets relative to a vertex cover of cardinality $C$.
Given a constant $\nu \,( 0<\nu \leq 1)$ and $C$, the non-empty set,
\begin{equation}\label{E:Lck}
\mathnormal{L}_{\nu,C}\triangleq\{A: A\subseteq V, |A|\leq C, \, r(A)\geq \nu\}
\end{equation}
 defines a class of optimal and near optimal subsets, with the degree of near optimality depending on $\nu$. Although we frequently observe that subsets of a vertex set or a particular optimal set are in $\mathnormal{L}_{\nu,C}$ that is not always the case.
In fact it may not contain any subsets of the vertex cover.
Let $m$ be the smallest cardinality of sets in $\mathnormal{L}_{\nu,C}$. Let $\mathbf{S}$ be any class of sets,  
 $\mathbf{S}\subseteq\{X \subseteq \mathnormal{L}_{\nu,C}, \, |X|=m \}$.  This is the class of starter sets for our method and it is determined by calculating $F$ exactly for all subsets of cardinality $m$. 

Using the definition (\ref{E:Lck}) it is not hard to show that if the set  $S \in \mathnormal{L}_{\nu,C}$, then  $S \cup \{u\} \in \mathnormal{L}_{\nu,C}$, (see e.g. Lemma 1 \cite{Hunt2}). Thus any superset of a near optimal set is also near optimal and the offered solution obtained by a greedy extension of $\mathbf{S}$ is near optimal with a rank of at least $\nu$. The degree of near optimality $\nu$, determines $m$ and therefore the cardinality of sets whose values of $F$ would need to be obtained by exact evaluation. To obtain a greater degree of near optimality, i.e. a large $\nu$, a larger $m$ is needed.  Conversely a fixed value of
$m$ determines $\nu$. To limit the amount of exact computation of $F$, we choose $m=1$ or $m=2$. This choice can also limit the degree of optimality of the starter set depending on the graph topology. For graphs with a tree like topology and large hubs however, the corresponding value of $\nu$ can be quite high even with these small choices of $m$.  If $S^{*}$ is the offered solution it was shown in  \cite{Hunt1},\cite{Hunt2} that it is at least as good as the greedy solution and the degree of improvement can be calculated from $S^{*}$ itself.
When more is known calculation of $S^{*}$ is not needed to estimate the degree of improvement. Suppose $S$  with $|S| \leq K$ is an extension of a set in $\mathbf{S}$, and the normalized rank $\rho(S)$ (see section \ref{S:quality}) satisfies $\rho(S)> \eta\geq \rho(S_{g})$. Then the degree of improvement has an explicit lower bound in terms of $\eta$. Convenient candidates for such an $S$ are subsets of the reference vertex cover. Our method reduces to the classic greedy algorithm when $m=1$ and the resulting starting set of singletons reduces to the optimal set for $K=1$.
\subsection{Algorithm Description and Examples} \label{S:Ex}

To solve the problem (\ref{E:opt}) for $K < C$, we choose a collection of sets $\mathbf{S} \subset L_{\nu,C}$. Each set in $\mathbf{S}$ has cardinality $m$, the minimum cardinality of sets in $L_{\nu,C}$. Note that the exact solution is an element of this class whenever $m \leq K \leq C$. The output of this method, i.e. the offered approximation is the best set that results from a greedy extension of each set in $\mathbf{S}$, to a set of cardinality $K$. Since exact calculation of $F$ occurs for sets up to cardinality $m$ only, the method has complexity $O(N^{m}E)$ where $E$ is the complexity of evaluating $F$ for a single subset. Therefore when $m=1$ the complexity of our method is that of the classic greedy algorithm but it is greater when $m=2$. In  \cite{Hunt2} we assumed that the value of $F$ was obtained by solving equation (\ref{E:lineq}) by matrix inversion of an $N X N$ matrix. In this case $E=O(N^{3})$. \\
The spectral sparsification theory and associated development of nearly linear time algorithms for solving certain classes of systems of equations with Laplace matrices the work of D.A. Spiegelman and S.H. Teng \cite{Spiegelman}, opens the possibility of a significantly reduced complexity for evaluating $F$ at a single set namely $E=O(N\log^2N)$. Thus the complexity of our method is then $O(N^2\log^2N)$ and $O(N^{3}\log^2N)$ for $m=1$ and $m=2$ respectively. The relation between the linear equation (\ref{E:lineq}) used to calculate $F$ and the systems treated by fast Laplacian solvers can be easily seen.
Recall that $\mathbf{L}$, the normalized Laplacian discussed by Spiegelman and Teng is, \\
\begin{equation}
\mathbf{L}=D^{-1/2}\mathbb{L}D^{-1/2} \\  \nonumber
\end{equation}
where $\mathbb{L}=D-\mathbb{A}$, $D$ is an $N\times  N$ diagonal matrix whose $i$th diagonal entry is $d_i$=deg($i$), and $\mathbb{A}$ is the adjacency matrix of the graph. For the uniform random walk
$I-P=D^{-1}\mathbb{L}$. Thus on substituting this relation into equation (\ref{E:lineq}) we obtain an equivalent linear equation involving $\mathbf{L}$.
\begin{equation} \label{E:fastLapl}
\mathbf{L}u=b  ,\\
\end{equation}
where $u=D^{1/2}h$ and $b=D^{1/2}\mathbf{1}$.

\begin{figure}[tbp] 
  \centering
  \includegraphics[width=6.63in,height=3.32in,keepaspectratio]{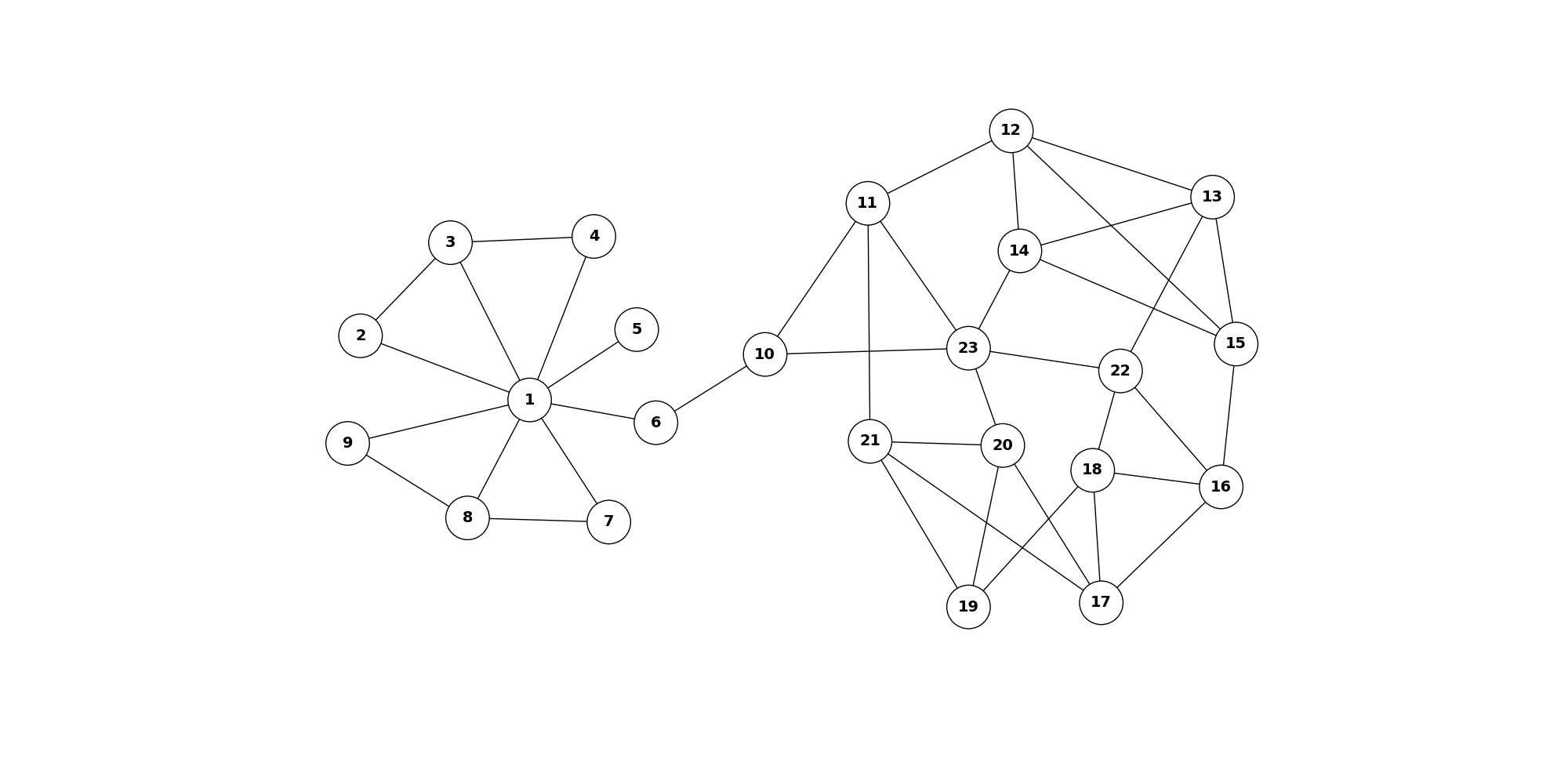}
  \caption{Graph with $N=23$ vertices, $E=97$ edges. The vertex cover has 12 elements (see text)}
  \label{fig:HuntFer1}
\end{figure}

We illustrate the method for solving problem (\ref{E:opt}) using the graph depicted in Figure \ref{fig:HuntFer1} that appears in a study of node centrality \cite{Duanbing}. It has 23 nodes graph and there is a vertex cover with 12 elements $\{1,3,8,10,12,13,15,16,18,20,21,23 \}$. 
Here we will show how our method generates offered solutions to the problem (\ref{E:opt}) for $1 \leq K \leq 12$. 
The minimum cardinality of sets in the class of near optimal sets $L_{.50,41}$  is $m=1$ so determining the exact solutions of equation (\ref{E:lineq}) is only needed for $K=1$.
Using the 15 one element sets of $L_{.50,12}$ as a starter set, all of the optimal sets up to $K=12$ are obtained by their greedy extension. The optimal sets for $K=10-12$ are greedy extensions of the optimal set for $K=9$. Moreover as shown in Table I, all of the optimal sets(shown in italics) can be obtained from the extension of just 4 elements in the starter set. 

\begin{figure}[tbp] 
  \centering
  \includegraphics[width=7.27in,height=4.85in,keepaspectratio]{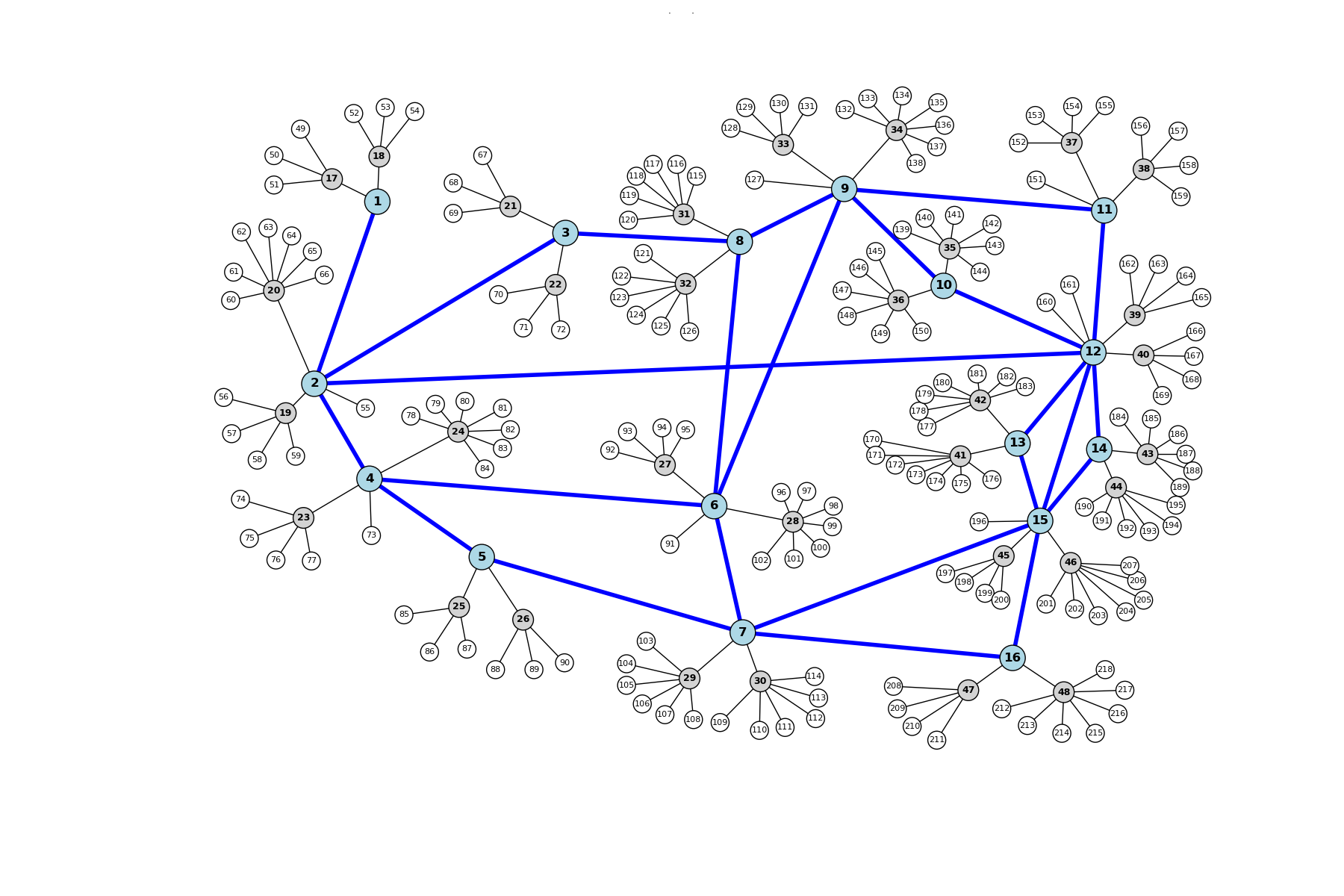}
  \caption{A compressed version of an actual IPS Network: N=218 vertices, E=226 edges. The vertex cover is $\{2,3,4,6,7,9,12,15,17-48\}$}
  \label{fig:HuntFer2}
\end{figure}

\begin{figure}[tbp] 
  \centering
  \includegraphics[width=7.38in,height=3.31in,keepaspectratio]{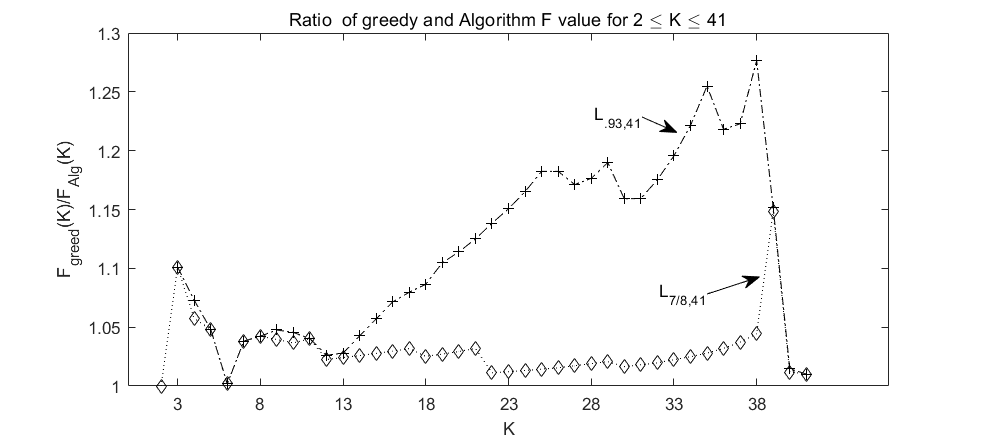}
  \caption{Comparison of $F$ values calculated by Algorithm with Greedy solution values for starter sets $L_{7/8,41}$ and $L_{.93,41}$ for the graph shown in Figure \ref{fig:HuntFer2}}
  \label{fig:HuntFer3}
\end{figure}

Figure \ref{fig:HuntFer2} is the graph model of a Internet Service Provider (ISP) network \cite{CDab}. It has 218 nodes and there is a vertex cover with 41 nodes (see figure). The optimal sets for $2\leq K \leq 41$ were approximated using the algorithm and Figure \ref{fig:HuntFer3} shows the results of using two different starter sets. For the near optimal
class $L_{7/8,41}$ , we have $m=1$,and for $m=2$, a second starter set comes from $L_{.93,41}$. The performance of each starter set is evaluated by calculating the ratio $\frac{F_{greed}(K)}{F_{alg}(K)}$ for each value of $K$. Here $F_{greed}(K)$ is the estimate of the optimal value of $F$ for cardinality $K$ obtained by the classical greedy method and $F_{alg}(K)$ is the estimate of the optimal value using the algorithm. As a consequence of Corrollary 1  in \cite{Hunt2}, this ratio is always at least $1$. The generally larger values of the ratio for $L_{.93,41}$ show the larger improvement over the greedy solution in comparison to $L_{7/8,41}$.


\begin{table*} \scriptsize
\caption{\small{Greedy extension of near optimal singletons of graph in Figure \ref{fig:HuntFer1}-optimal sets in italics}}  
\begin{center}
%
\begin{tabular} {| l | l | l | l | l | p{.86in} |}
\hline
K=1    &     2       &  3       &  4          &  5  &  6       \\ \hline

$\mathit{23}$  &$\mathit{ 23, 1 }$   &$\mathit{23,1,16}$  & 23,1,16,21 &$\mathit{23,1,16,21,12}$  & \\   \hline

17  & 17,1     & 17,1,23&$\mathit{ 17,1, 23,13}$         &       &       \\ \hline

13  & 13,1          &13, 1, 20  &              &      &        \\  \hline

18  &18, 1  &18,1,23    &18,1,23,12  &18,1,23,12,    &$\mathit{18,1,23,12,15}$ \\  \hline
\end{tabular}
\end{center}
%
%
%
\begin{center}
\begin{tabular} {|l| l| l| l|} 
\hline
K=7 &   8 &    9  \\ \hline

    &        &$\mathit{23,1,16,21,12,18,3,8,13}$ \\ \hline
    &           &      \\  \hline
    &           &        \\ \hline
$\mathit{18,1,23,12,21,15,3}$ & $\mathit{18,1,23,12,21,15,3,8}$ &      \\ \hline
\end{tabular}
\end{center}
\end{table*}
%
%
%

%

\subsection{Quality of the approximation}\label{S:quality}

Following Ilev (\cite{Ilev}) when $F$  is supermodular, the value for the the empty set can be defined as 
\begin{equation}
0\leq F(\emptyset)=\max_{X \cap Y=\emptyset , X, Y \subseteq V} F(X) +F(Y)-F(X \cup Y) < \infty
\end{equation}
Thus by the definition of $r \, \,$,  $r(\emptyset)=\frac{F_{max}-F(\emptyset)}{F_{max}-F_{min}}$.
This means the function $\rho$ defined on sets $A\,$ by $\rho(A)=r(A)-r(\emptyset)$ is bounded, submodular and non-decreasing and satisfies $\rho(\emptyset)=0$. Also note that since $F(\emptyset) \geq F_{max}$, it follows that $\rho(A) \geq 0$ for all $A \subseteq V$. Following the nomenclature of \cite{Wang}, $\rho$ is called a normalized rank function. Since $\rho$ is an affine function of $F$, the optimization problem (\ref{E:opt}), is equivalent to the problem of finding the set that  maximizes the normalized rank subject to the same constraints. This is a special case of the general problem first considered by Nemhauser, Wolsey and Fisher. Using their result (\cite{NemWolFish}, Section 4) we showed that the ratio of the normalized ranks of the classic greedy solution and the optimal solution has a lower bound of $(1-\frac{1}{e})$ \cite{Hunt2}.
Let us now assume that the $m$ stage greedy solution is in $\mathbf{S}$. 
This is not a serious restriction for when $S_{g} \notin {S}$, it can always be added to $\mathbf{S}$. \\
\begin{proposition}\label{P:bound}
If $S^{*}$ is the offered solution constructed by the method described in sections \ref{S:optnopt}, \ref{S:Ex}, then 
\begin{equation}
\rho(S^{*})\geq \left (1-\frac{1}{e} \right)\rho\left(\mathcal{O}_{K}^{*}\right)
\end{equation}
where $\mathcal{O}^{*}_{K}$ is an optimal set of cardinality $K$.\\
\end{proposition}
The proof uses the fact that the greedy solution $S_{g}$ of cardinality $K$ satisfies the inequality $\rho(S_{g})\geq \left(1-\frac{1}{e}\right)\rho(\mathcal{O}^{*}_{K})$. \\
Once $F(S^{*})$ and $F(S_{g})$ are computed we can determine $\chi$ such that $\rho(S^{*})=(1+\chi)\rho(S_{g})$. Thus the quantity $\chi$ measures the degree of improvement of our method over the greedy approach. Suppose $F(S^{*})$ has not been computed? If $F(S_{g})$  is known, then an estimate of $\chi$ can easily be obtained in the course of computing $S^{*}$.  To see this, suppose  there is an $S \subset V,\, |S| \leq K$, for which $F(S)<F(S_{g})$. Then $\rho(S) \geq \eta > \rho(S_{g})$ for some $\eta > 0$. If $S$ is a greedy extension of the a set in $\mathbf{S}$, then  $\rho(S^{*}) > \rho(S_{g})$. The degree of improvement of $S^{*}$ over $S_{g}$ has a lower bound on $\chi$ that involves $\eta$. For example if we choose $\delta=1-\frac{\rho(S_{g})}{\eta}$ 
then $\chi > \frac{\delta}{1-\delta}$. \\

A $(1-\frac{1}{e})$ lower bound similar to Proposition \ref{P:bound}  was established by Borkar et al in \cite{Borkar}. 
Specifically it is a lower  bound on the ratio of $F(S_g(a))-F(\{a\})$ 
to $F(\mathcal{O}^{*}_{k})-F(\{a\})$, where $S_g(a)$ is the result of the greedy algorithm starting with singleton $a$. \\
%
\section{ An upper bound and a surrogate for $F$} \label{S: Fupbound}
In this section we derive upper bounds for $F(A)$ in terms of some topological aspects of graphs induced by $A$ and $V\setminus A$.
As a byproduct we also derive a surrogate set function from the general upper bound of section \ref{S:Fupperbound} that can be used to select desirable candidates for the solution of problem (\ref{E:opt}) without directly computing $F$. 
 
We first assume that $A$ belongs to the class of one dominant or dominating sets,
and then obtain an upper bound on $F$. In section \ref{S:Fupperbound}, an upper bound for arbitrary sets is obtained based on a recent first hitting time formula obtained by Basu et al \cite{Basu}.
 Both approaches use a representation of $\mathbb{E}_{i}[T_A]$ for a  random walk starting at $i \notin A$ in terms of the expected number of times a directed edge of the form $e=(j,k)$, $j , \, k \notin A$ is visited plus the expected number of times an edge of the form $f=(j,k)$, $j \notin A, k \in A$ is visited. Since an edge ending in $A$ can only be visited once, and since this occurs with probability one, the expected number of times such an edge is visited is $1$. 

Let $n_{e}$ be the number of times a random walker starting at $i \notin A$ crosses an edge $e$ before $T_A$. 
For a non-lazy random walk with no self-loops (for example the uniform random walk on $\mathcal{G}$), 
the elapsed time between the start of a walk at $i \notin A$, and the time of first arrival in $A$ is the sum of unit time steps where the walker either moves along an edge joining $A^{c}$ to $A$ and is absorbed (the expected number of such steps being 1) or moves along an edge $e$ that has both endpoints in $A^{c}$. Such an edge is called uncovered. Thus,
\begin{equation} \label{E: pathrep1}
T_{A}=1 +\sum_{n=0}^{\infty}\{\sum_{e~ uncovered}\mathbf{1}_{e}(X_n,X_{n+1})\mathbf{1}(n < T_{A})\}.
\end{equation}
Thus the expected value of $T_{A}$, starting at $i$ can be written as,
\begin{equation} \label{E:timeeq2}
\mathbb{E}_{i}[T_{A}]=1+\mathbb{E}_{i}[\sum_{n=0}^{T_{A}-1}(\sum_{e~ uncovered}\mathbf{1}_{e}(X_{n},X_{n+1}))]. \\
\end{equation}
Now equation (\ref{E:timeeq2}) can be rewritten as,
\begin{equation}  \label{E:Ioisef}
\mathbb{E}_{i}[T_A]=1+\mathbb{E}_{i}[\sum_{\,\, e: \textit{uncovered}}n_{e}]. \\
\end{equation} \\
\textbf{REMARK:} Uncovered edges can be divided into edges comprising the connected components of $V \setminus A$. So
for a fixed node $i$,
equation (\ref{E:Ioisef}) can be rewritten as \\
\begin{equation} \label{E:Ioisef2}
 \mathbb{E}_{i}[T_A]=1+\mathbb{E}_{i}[\sum_{\,\, e: e\in \mathcal{T}_{i}}n_{e}], \\
\end{equation}
where $\mathcal{T}_{i}$ is the component containing $i$. \\
\subsection{Upper bound on $F$ for one dominant sets} \label{S: onedom}

If $A \subset V$, and every element of $V\setminus A$ is adjacent to some element of $A$, the set is called one dominant or dominating.  $A$ is said to be independent if none of its elements are adjacent to each other. Dominating sets, particularly those of small cardinality have been extensively used in the design of wireless and sensor networks. Moreover the role of minimum cardinality one dominant sets in network controllability and in the regulation of protein-protein interaction networks is the subject of intensive study as we discussed in the introduction. Using ( \ref{E:Ioisef}), we can derive a simple upper bound on $F$ that can be useful in identifying higher quality target sets without calculating $F$. If $e=(j,k)$ it is known that $\mathbb{E}_{i}[n_{e}]=\gamma^{i}(j)p(j, k)$, where $\gamma^{i}(j)$ is the expected number of times a walker starting at node $i$ visits $j$ before $T_A$ (p.74 \cite{Pitman} ). Thus we have,
\begin{equation}\nonumber
\mathbb{E}_{i}[T_A]=1+\sum_{j \notin A}\left[\sum_{k: k\notin A, (j,k) \in \mathcal{E}}\gamma^{i}(j)p(j, k)\right], \\
\end{equation}
which implies the inequality,
\begin{equation} \label{E:sigmaeq}
\mathbb{E}_{i}[T_A]\leq 1 + S(i)\sigma^{*}
\end{equation}
where $\sigma^{*}=\max_{j \notin A}(\sum_{k \notin A}p(j,k))$ and $S(i)=\sum_{j \notin A}\gamma^{i}(j)$.
We note using $Q=P_{A}$, the standard notation for absorbing Markov chains , that $\gamma^{i}(j)=(I-Q)^{-1}_{ij}$.
Since $S(i) \leq \mathbb{E}_{i}[T_A]$, 
%
%
We can therefore conclude that
\begin{lemma}
Let $A$ be a one dominant subset of $V$. Then $F(A)$ has the upper bound,
\begin{equation} \label{E:DomUB}
F(A) \leq \frac{N-|A|}{1-\sigma^{*}}
\end{equation}
\end{lemma}
\textbf{Proof:}  Summing inequality (\ref{E:sigmaeq}) over all $i \in  A^{c}$ yields the inequality 
\begin{equation} 
F(A) \leq N-|A| + \left(\sum_{i \in A^{c}}S(i) \right)\sigma^{*}\leq N-|A| + F(A)\sigma^{*}
\end{equation} \\
Since $A$ is one dominant $\sigma^{*} < 1$ and hence inequality (\ref{E:DomUB}) is an immediate conclusion. $\Box$  \\ \\
Equality holds when $\sigma^{*}=0$, i.e. when $A^{c}$ is independent. In this case $A$ is a vertex cover. \\ \\
\textbf{Remark:}
Since the probability of entering $A$ from a node $i \notin A$ is a least $1-\sigma^{*}$, it is natural to suspect that 
if $\sigma^{*}$ is small, $A$ will be near optimal in the sense of section \ref{S:optnopt}. In fact if $C$ is the cardinality of a vertex cover and $|A|< C$, then $A \in L_{\nu^{*},C}$ where $\nu^{*}=1-\frac{\frac{N-|A|}{(1-\sigma^{*})}-(N-C)}{F_{max}-(N-C)}$.

%
NOTE: Let $\sigma(j)=\sum_{k \notin A}p(j,k)$. For the uniform random walk, $\sigma(j) \leq 1-\frac{1}{deg(j)}$ so $\sigma^{*}$
in equation (\ref{E:sigmaeq}) satisfies $1-\sigma^{*} \geq \frac{1}{d^{*}}$ where $d^{*}$ is the maximum degree (equality holds when $\mathcal{G}$ is a star with a central node of degree $d^{*}$). It follows
that for the uniform walk we have the upper bound
\begin{equation}
F(A)\leq (N-|A|)d^{*}
\end{equation}
%
\subsection{Upperbound on $F(A)$ for general $A$} \label{S:Fupperbound}
To obtain an upper bound on $F$ for a general set, we will use a recent result of Basu et al (\cite{Basu}) from the mixing theory of Markov chains. We assume the chain is irreducible and reversible therefore there exists a unique stationary distribution for
the chain, $\pi$. Basu et al discovered a formula for the expected first hitting time to a set $A$ starting from an initial distribution of points in $A^{c}$ on the boundary of $A$ that were in $A$ at the previous time step. The expected first hitting time is equal to the bottle neck ratio (sometimes called the conductance). It is used to define  Cheeger's constant, a lower bound on the mixing time of the chain (\cite{Levin}).


The bottle neck ratio \cite{Levin} of set $B$, $\Phi(B)$ is defined as,
\begin{equation} 
\Phi(B)=\frac{\sum_{x \in B, \, y \in B^{c}}\pi(x)P(x,y)}{\sum_{x \in B}\pi(x)}
\end{equation}
For the special case of the uniform random walk this is, \\
\begin{equation}
\Phi(B)=\frac{|\partial B|}{\sum_{b \in B}deg(b)}
\end{equation}
Here $|\partial B|$ is the number of edges connecting $B=A^{c}$ to $A$.\\
%
%

  Basu et al obtained an expression for the expected first hitting time to a set $A\subset \Omega$, for a random walk starting at a node in $A^{c}$ adjacent to $A$. Specifically consider the possible destinations in $B=A^{c}$ of a walk that at the previous time step was in $A$. They define a probability distribution on the possible 
landing points in $A^{c}$, $\Psi_{A^{c}}(y)=P_{\pi_{A}}[X_{1}=y | X_{1} \in A^{c}]$. Here $\pi_{A}$ is the stationary probability distribution restricted to $A$. 
\begin{equation}
\pi_{A}(x)=\frac{\pi(x)\mathbf{1}_{A}(x)}{\pi(A)} \nonumber
\end{equation}
For an irreducible, reversible chain
the expected first hitting time to $A$ given the initial distribution $\Psi_{A^{c}}$, is expressed in terms of the bottleneck ratio.
\begin{theorem} \cite{Basu}  \label{T: Basu}
Let ( $\Omega$, $\mathbb{P}$, $\pi$ ) be a finite irreducible reversible Markov chain.  Let $A \subsetneq V$ with complement $B$ be given. 
Then,
\begin{equation}
\mathbb{E}_{\Psi_{B}}[T_{A}]=\frac{1}{\Phi(B)} .
\end{equation} 
\end{theorem}
Suppose a random walk begins at $b$ in the support of $\Psi_{B}$, and then visits a node $i \notin A$ before reaching $A$ for the first time. By the Strong Markov property, the number of times an uncovered edge $e$ is crossed by the walk after the first visit 
to $i$ but before $T_A$, is the number of times $e$ is crossed when  starting at $i$ before
 $T_A$. Let $T^{y}_{i}$ for $y \notin A$  be the time for a random walker starting initially at $y$,
to arrive at $i$ for the first time. If $i$ is a non-isolated point of $A^{c}$, and $y \in \mathcal{T}_{i}$, then such a path exists with positive probability. The hypotheses on the Markov chain are the same as those in Theorem \ref{T: Basu}.
\begin {lemma} \label{L:rwpath1}
Let $i\in B=A^{c}$, be a non-isolated point in $B$ and suppose $y$ is adjacent to $A$.  Then
\begin{itemize}
  \item $y \in supp(\Psi_{B})$
  \item If $y \in \mathcal{T}_{i}$ (the connected component of $V\setminus A$ that contains $i$
   then $T^{y}_{i} < T^{y}_{A}$ with positive probability \\
\end{itemize}  
\end{lemma} 
\textbf{Proof:} 
Since $y$ is adjacent to $A$ there is an edge $(a,y)$ with $a \in A$, thus $p(a,y)>0$. Using the definition of $\Psi_{B}$, one can write,
\begin{equation} \label{E:psieq}
\Psi_{B}(y)=\frac{\sum_{a \in  A}\pi(a)p(a,y)}{\sum_{b \in B,\, a\in A}\pi(a)p(a,b)}
\end{equation}
Thus, $\Psi_{B}(y)>0$ since $\pi(a)>0$ for all $a$. To show that $T^{y}_{i} < T^{y}_{A}$ with positive probability, note that when $i=y$, then the inequality holds with probability 1. If $i \ne y$, then the connectedness of $\mathcal{T}_{i}$ and the reversibility of the Markov chain imply the existence of a path from $y$ to $i$ that lies entirely 
in $A^{c}$. Thus there is a sample path starting at $y$ and ending at $i$ and since the transition probability on any edge of the path is positive, the sample path has positive probability by the Chapman Kolmogorov equality. For such a path $T^{y}_{A}=T^{y}_{i}+T^{i}_{A}$ so the claimed inequality holds with positive probability.
$\Box$. \\ \\
\textbf{REMARK:} Let $\delta A$ be the nodes in $A^{c}$ that are adjacent to $A$. The lemma implies that for fixed non-isolated
 $i \notin A$, any $b \in \delta A \cap \mathcal{T}_{i}$ is in the support of $\Psi_B$. If $i$ is isolated then it is automatically in $supp(\Psi_{B})$ and we have $y=i$ that is, $T^{y}_{i}=T^{i}_{i}< T^{y}_{A}$ with probability one.  \\ \\

We now turn to the derivation of an upper bound for $\mathbb{E}_{i}[n_{e}]$. The upper bound accounts for the likelihood that $i \notin A$ is visited by a random walk started at $b$ before the walk hits $A$ for the first time.
\begin{proposition} \label{P: istartbound}
\begin{equation} \label{E:iacess} 
\mathbb{E}_{i}[n_{e}]\leq
\frac{\mathbb{E}_{\Psi_B}[n_{e}]}{\sum_{b \in \delta A}\Psi_{B}(b)\mathbb{P}(T^{b}_{i} < T^{b}_{A})} \\ \\
\end{equation}
\end{proposition}
\textbf{Proof:} Let $e$ be an uncovered edge in the component $\mathcal{T}_{i}$. We define $n_{e}^{i}$ to be the number of times a random walk path crosses an uncovered edge $e$ during the time period, $T_i\leq t \leq T_A$ with $n_{e}^{i}=0$ if $T_{i} > T_{A}$.
It can be seen that, for $b \in supp(\Psi_{B})$, 
\begin{equation}
\begin{aligned}
\mathbb{E}_{b}[n_{e}] \geq \mathbb{E}_{b}[n^{i}_{e}]=\mathbb{E}_{b}[n_{e}^{i}; T_{i}^{b} < T_{A}^{b} ]= \\  
\mathbb{E}_{b}[n_{e}^{i} \,| T_{i}^{b} < T_{A}^{b}] \cdot\mathbb{P}[T_{i}^{b} < T_{A}^{b}]. \\
\end{aligned}
\end{equation}
By the Strong Markov property, $\mathbb{E}_{b}[n_{e}^{i} \, \, | T_{i}^{b} < T_{A}^{b}]=\mathbb{E}_{i}[n_{e}]$.
Thus 
\begin{equation} \label{E: stmarkovineq}
\mathbb{E}_{b}[n_{e}] \geq \mathbb{E}_{i}[n_{e}]\cdot \mathbb{P}[T_{i}^{b} < T_{A}^{b}]. \\
\end{equation} 
Multiplying each side of inequality (\ref{E: stmarkovineq}) by $\Psi_{B}(b)$ and adding over all $b \in supp(\Psi_{B})$ results in the inequality
\begin{equation} \label{E: psineq}
\mathbb{E}_{\Psi_B}[n_{e}]\geq \left(\sum_{b \in supp(\Psi_B)}\Psi_{B}(b)\mathbb{P}[T_{i}^{b} < T_{A}^{b}] \right)\mathbb{E}_{i}[n_e]
\end{equation}
By Lemma \ref{L:rwpath1} and the remark that follows its proof, the sum on the right hand side of (\ref{E: psineq})
is non-zero and the set $supp(\Psi_B)$ can be replaced by $\delta A$. Thus the statement of the proposition follows from (\ref{E: psineq}). $\Box$  \\

An upper bound on $\mathbb{E}_{i}[T_{A}]$ can be deduced from Proposition \ref{P: istartbound}. First we have
for any uncovered edge $e$,  $\mathbb{E}_{\Psi_B}[n_e] \leq \mathbb{E}_{\Psi_B}[T_A]$. Thus from equation (\ref{E:Ioisef}) it follows
that,
\begin{equation} \label{E: TAupperbound}
\mathbb{E}_{i}[T_A]\leq 1+ \mathbb{E}_{\Psi_B}[T_A]\left(\sum_{e \in \mathcal{T}_{i}}p_{e}\right)
\left(\frac {1}{d(i,A)} \right) \\
\end{equation}
where $p_{e}=1$ if $e$ is an uncovered edge and is $0$ if $e$ is covered,
and $d(i,A)= \sum_{b \in \delta A}\Psi_{B}(b)\mathbb{P}(T^{b}_{i} < T^{b}_{A})$.  \\
%
The upper bound on $F(A)$ then follows from inequality (\ref{E: TAupperbound}).
\begin{corollary}\label{C: FAupperbound} For $B=A^{c}$,
\begin{equation} \label{E: FAineq}
F(A) \leq  N-|A| + \frac{1}{\Phi(B)}\left(\sum_{e \in \mathcal{E}}p_{e}\right )
\left( \sum_{i \notin A}\frac {1}{d(i,A)}\right ). \\ \\
\end{equation}
\end{corollary} 
%

%
\textbf{REMARK:} It is interesting to note that nothing about the connectivity $V\setminus A$ was assumed in the derivation of the upper bound in (\ref{E: FAineq}). Equality holds when $A$ is a vertex cover since in this case $\sum_{e \in \mathcal{E}}p(e)=0$. \\ \\

The inequality (\ref{E: FAineq}) has two useful consequences. First, it gives insight into the conditions for fast consensus. Like the work of Pirani (\cite{Pirani2}), Pirani and Sunderam (\cite{Pirani3}), Xia and Cao (\cite{Xia}) on bounds for the smallest eigenvalue of grounded Laplacians, the bound depends on graph theoretic properties of the (leader) stubborn nodes and the connectivity of the residual graph of (follower) non-stubborn nodes. Secondly in the uniform random walk case there is a simplified form of (\ref{E: FAineq}) that is an effective tool for identifying highly optimal sets without actual computation of $F$ (section \ref{S: surrogate}). 
When all the candidate sets have the same cardinality, using the surrogate to select nearly optimal K element sets. This is a natural way to improve the offered approximation of the greed extension algorithm.

The upper bound in Corollary \ref{C: FAupperbound} depends on three somewhat independent graph properties of $A$ and $V \setminus A$.
The bottle neck ratio (a consequence of Theorem 1) counts the fraction of edges joining $A$ and $V\setminus A$, and the second the number of uncovered edges, depends of the ability of $A$ to minimize the connectivity of the residual set $V \setminus A$. In this respect, an optimal or near optimal $A$ resembles a near optimal solution of the partial vector problem (\cite{Caskurlu}) or critical node problem (\cite{Arulselvan}); important discrete optimization problems with applications in network security and communication.
The third factor is node specific. 
The quantity $d ( i, A)=\sum_{b \in \delta A}\Psi_{B}(b)\mathbb{P}[T_{i}^{b} < T_{A}^{b}]$, measures the accessibility of the node $i \notin A$, to $A$. It is the weighted probability that $i$ can be reached by a random walker starting at a node $b$ that is adjacent to $A$ before the walker visits $A$ for the first time. Nodes $i$ that are remote have a small $d(i,A)$ value resulting in a large contribution to the bound.

The accessibility quantity is very closely related to a centrality parameter 
that arises in the study of opinion dynamics in a social network with     
stubborn agents \cite{Yildiz}. The context is a binary voter model with
rival sets of stubborn agents holding different opinions (labeled 0 and 1 
respectively). The authors define the quantity $r(i, A)=\sum_{j \notin V_0 \cup V_1}\mathbb{P}[T^{j}_{i} < T^{j}_{A}]$ , where $A=V_{0}\cup V_{1}$ and  $V_{0}$, $V_{1}$ are the nodes holding opinions $0$ and $1$ respectively. Thus, $r(i , A)$  measures how influential $i$ is in shaping the opinions of other non-stubborn nodes in the presence of stubborn agents in $V_{0}$ and $V_{1}$  (page 19.13, page 19.9 \cite{Yildiz}). In the presence of a single fixed set of stubborn agents holding opinion $0$, $r(i,A)$ is an unweighted version of $d(i,A)$. In the optimal placement problem the authors consider the best location for a rival stubborn node $i$ 
holding opinion $1$. This reduces the problem to maximizing an objective function (the probability of absorption by $i$ before absorption by the set $A$).  The optimal placement is treated in a manner similar to what we do here. However we do not use a gluing argument as in \cite{Yildiz}, thus we can analyze the case where $V\setminus A$ has multiple components. This can be useful when the number of connected components of $V\setminus A$ is known for example when starter sets are good approximations of the solution of the critical node problem \cite{Arulselvan}.

 There are connections between the $d(i,A)$ and the eigenvector components of the zero eigenvalue of the graph Laplacian  \cite{GhaderiSrikant}, \cite{Pirani2}, \cite{Xia}. This connection merits discussion because the eigenvalue approach may lead to methods of altering the graph topology to facilliate convergence and other desirable network behavior. In the next section, the theory of absorbing Markov chains is used to explain the link between $\{d(i,A): i \in A^{c}\}$ and the eigenvectors of the graph Laplacian matrix. The use of absorbing Markov chains in consensus models is not new. Pirani used the theory to derive upper and lower bounds on the maximum eigenvalue of the inverse of the grounded Laplacian associated with $A$. Researchers 
(see e.g. \cite{Acemoglu}, \cite{GhaderiSrikant}, \cite{Pirani2},\cite{Pirani}, \cite{Como}, \cite{Yildiz}) have also used the theory to derive  representations for the consensus value itself.

\subsection{Connections with the Graph Laplacian} \label{S: Lapl}
Let $P$ be the transition matrix of a Markov chain with $n$ absorbing
states corresponding to nodes $\mathcal{S}=1:n$. The graph of the Markov chain is a directed graph with no outgoing edges at nodes $\mathcal{S}$. The Laplacian matrix for the graph is $\mathcal{L}=I-P$. After appropriate labeling of nodes $P$ can be put in  the standard form for absorbing chains (\cite{Kemeny})
\begin{equation} \label{E: Absorb}
P=
\left (
\begin{matrix}
I & O \\
R&Q
\end{matrix}
\right)
\end{equation}
where $n=|\mathcal{S}|$,  $I$ is an $n \times n$ identity matrix, $O$ is a $n \times (N-n)$ matrix of zeros, $R$ is an $(N-n) \times n$
matrix and $Q$ is an $(N-n) \times (N-n)$ matrix. The vector $x \in \mathcal{R}^{N}$ is an eigenvector of $\mathcal{L}$ corresponding to the eigenvalue $\lambda_{\mathcal{L}}=0$ if and only if it is an eigenvector of $P$ belonging to $\lambda_{P}=1$.
The algebraic multiplicity of $\lambda_{\mathcal{L}}$ is $n$ and it is semisimple (\cite{Agaev},\cite{Chebot1}). Therefore the eigenspace is spanned by $n$ independent vector solutions of $\mathcal{L}x=\mathbf{0}$.  We now turn to a discussion of how these components are very closely related to the upper bound of $F$ in Corollary \ref{C: FAupperbound}. Using equation (\ref{E: Absorb}), and dividing the vector $x$ into components corresponding to absorbing and non-absorbing states $x=\left ( \begin{matrix} \hat{x} \\ \nu \end{matrix} \right)$ one obtains,
\begin{equation} \label{E: Beq}
(I-Q)\nu=R\hat{x}.
\end{equation}
Here $I$ is an $(N-n) \times (N-n)$ matrix, $\nu \in \mathcal{R}^{N-n}$ and $\hat{x} \in \mathcal{R}^{n}$. The eigenspace of $\lambda_{\mathcal{L}}=0$ is spanned by vectors $x_{i}=\left ( \begin{matrix} \hat{e}_{i} \\ \nu_{i} \end{matrix} \right), \, i=1, \cdots n$.
For each $i$ the vector $\hat{e}_{i}$ is the vector with component $1$ in the $i$th position and $0$ elsewhere. It represents the absorbing node $i$ and $\nu_{i}$ is the solution of (\ref{E: Beq}) when $\hat{x}=\hat{e}_{i}$. Since $x_i$ is also an eigenvector of
 $P$ for eigenvalue 1, the matrix $\varLambda=\left[ \begin{matrix} x_1 & \cdots & x_n\end{matrix} \right]$ satisfies 
$P\varLambda=\varLambda$. The submatrix consisting of the last $N-n$ rows of 
$\varLambda$ is
$H=\left[ \begin{matrix} \nu_1 & \cdots & \nu_n \end{matrix} \right]$. This matrix has a probabilistic interpretation that is relevant to Corollary  \ref{C: FAupperbound}.
Let $s$ be a state in $\mathcal{S}$ and let $b \notin \mathcal{S}$. Then $H_{b,s}=\mathbb{P}[T_{s} < T_{\mathcal{S}\setminus s}|\, X(0)=b]$.
Therefore on setting $i=s$ and $A=\mathcal{S}\setminus s$ we see that $H_{b,i}=\mathbb{P}[T_{i}^{b}<T_{A}^{b}]$ and
$d(i,A)= \sum_{b \in \delta A}\Psi_{B}(b)H_{b,i}$.

The elements of the matrix $H$ are also weights used to represent the consensus value as a convex combination
of the initial values of the stubborn agents (see \cite{Como},\cite{Pirani2},\cite{GhaderiSrikant}).  To obtain a probabilistic interpretation of $d(i,A)$, say that $b \notin \mathcal{S}$ is attracted to $i \in \mathcal{S}$ if $i$ is the first node in $\mathcal{S}$ encountered in a random walk starting at $b$. Then $d(i,A)$ measures the attractiveness of $i$ to points adjacent to $A$.\\ \\
\textbf{REMARK:} $(I-Q)$ is the grounded Laplacian $L_{g}$ (\cite{Xia},\cite{Pirani}) of the Markov chain graph.
\subsection{A surrogate for $F$} \label{S: surrogate}
\begin{figure}[tbp] 
  \centering
  \includegraphics[width=5.87in,height=4.41in,keepaspectratio]{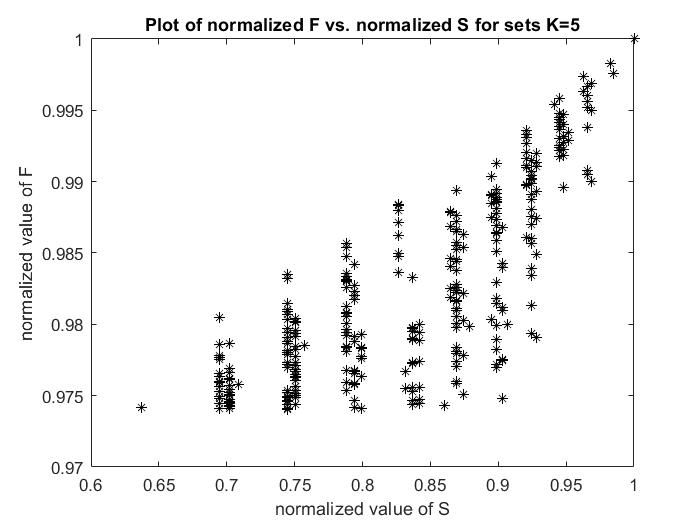}
  \caption{Plot of normalized values of F versus normalized surrogate value S. Values for top 8000 (of 33,649) K=5 sets of graph in Figure \ref{fig:HuntFer1} are plotted every 25 horizontal steps for readibility}
  \label{fig:HuntFer4}
\end{figure}

The lesson of section \ref{S:optnopt} is that in contrast to the classical method, the greedy extension of a collection of small cardinality sets of high quality produces improved approximations of optimal sets of specified cardinality $K$. However there is no provision for finding better approximations if all of the sets have the same cardinality $K$. We must rely on trial and error swapping of single elements followed by evaluation of $F$. We developed a set valued function $S$ based on a simplification of the upper bound in Corollary \ref{C: FAupperbound}, that approximately (but satisfactorily) identifies highly optimal sets from among a collection of sets of the same size. Thus the use of $S$ as an alternative to direct calculation of $F$ is the basis of a promising heuristic for  problem (\ref{E:opt}). The question we address in this section is how well can highly optimal sets be identified knowing only the number of uncovered edges, the degree of each node in $V \setminus A$ and the number of cut edges $|\partial A|$ ? The $d(i,A)$ terms are ignored , thus the influence of inacessible nodes is left out. Indeed obtaining this information would involve $a$ computation comparable to calculating $F(A)$. Thus ranking sets by $S$ is only a rough approximation of a ranking by $F$. Given the set $A$ and assuming the uniform random walk the surrogate $S(A)$ is,
\begin{equation}\label{E: surrogate}
S(A)=1+\left(\frac{\sum_{b \in A^{c}}deg(b)}{|\partial A|}\right)\sum_{e \in \mathcal{E}}p_{e}
\end{equation}
The quality of sets of the same cardinality can be expressed in terms of a normalized F value $\rho_{F}(A)=\frac{F_{max}-F(A)}{F_{max}-F_{min}}$ where $F_{max}=\max_{|A| \leq K}F(A)$ and $F_{min}=\min_{|A| \leq K}F(A)$. An analogous normalized $S$ can be defined, $\rho_{S}(A)=\frac{S_{max}-S(A)}{S_{max}-S_{min}}$. The normalized values can be used to illustrate the correlation between highly optimal sets with respect to the objective function $F$ and highly optimal sets with respect to the surrogate $S$.
Figure \ref{fig:HuntFer4}  shows a plot of $\rho_{F}$ versus $\rho_{S}$  for the graph shown in
Figure \ref{fig:HuntFer1} for a collection of $33,649$ $5$-element sets. 
%
%


\section{Conclusion} \label{S: Conclusion}
We considered the problem of maximizing the rate of convergence for the spread of consensus in the presence of stubborn agents in a network of fixed topology. We posed the optimization problem :determine the set of nodes $A$ of fixed maximum cardinality that leads to the fastest rate of convergence to consensus.  Here, following the random walk approach of \cite{Borkar}, \cite{Clark} the desired set
minimizes $F(A)$ , the sum of the first hitting times of random walks starting at a node outside the set $A$. In this paper we demonstrate that this choice of objective function has advantages that firstly enable the development of a potentially useful method of finding or approximating an optimal set and secondly allows us to gain some understanding of its structure. Using the facts that $F$ is supermodular and that a vertex cover of the graph $\mathcal{G}$ is always a solution of the optimization problem for its cardinality, we developed a generalization of the classic greedy algorithm by constructing a class of optimal and near optimal small size starter sets whose degree of optimality is defined in terms of a vertex cover. Previously a submodular function derived from $F$ was used to prove that the performance ratio of this algorithm is at least $(1-1/e)$ (Proposition \ref{P:bound}) and in this paper we provide sufficient conditions for our offered approximation to be an improvement. Indeed implementation on two graphs (see Figures \ref{fig:HuntFer1} and \ref{fig:HuntFer2}) demonstrate that the theoretical bound can be exceeded.The method is polynomial in $N$,the number of nodes and works well for moderate sized networks, but for very large networks the required computations of $F$ become onerous. In section \ref{S: surrogate}, we discussed a potential remedy for this. We presented heuristic evidence that an easy to calculate surrogate set function derived from an upper bound on $F$, can be used to identify highly optimal sets without direct computation of $F$, thus making it computationally cheaper to identify optimal and near optimal sets.

The use of first hitting time as a time to consensus provides some insight into the topological characteristics of optimal and near optimal sets in ways that are analogous to the earlier results of \cite{GhaderiSrikant} and \cite{Pirani}. The setting is an reversible irreducible random walk on $\mathcal{G}$, with transition matrix determined by the consensus model. A recent equality of Basu et al \cite{Basu} was used to derive an upper bound for $F(A)$ in terms of the bottle neck ratio of $A^{c}$ , the number of edges in $A^{c}$ non-adjacent to $A$ (i.e. the number of uncovered edges) and quantities that measure the accessibility of a node $i \notin A$ to a random walk started at a node adjacent to $A$. In fact we show (see section \ref{S: Lapl}) that these accessibility measures can be written in terms of components of the eigenvector associated with the smallest eigenvalue of an augmented Laplacian.  The work of \cite{Como} shows that the probabilities in equation( \ref{E: TAupperbound}) appear as weights in the representation of the consensus value.

\end{document}